\documentclass[]{spie}  

\newcommand{\unit}[1]{\ensuremath{\, \mathrm{#1}}}
\newcommand\arcsec{\mbox{$^{\prime\prime}$}}%
\newcommand\arcmin{\mbox{$^\prime$}}%
\usepackage{amsmath,amsfonts,amssymb}
\usepackage{graphicx}
\usepackage[colorlinks=true, allcolors=blue]{hyperref}
\usepackage{float}
\title{Extreme precision photometry from the ground with beam-shaping diffusers for K2, TESS and beyond}
\author[a,b,c]{Gudmundur Stefansson}
\author[a,b]{Suvrath Mahadevan}
\author[d]{John Wisniewski}
\author[a]{Yiting Li}
\author[a]{Marissa Maney}
\author[e]{Leslie Hebb}
\author[f]{Brett Morris}
\author[g,h]{Samuel Halverson}
\author[a]{Andrew Monson}
\author[i]{Paul Robertson}

\affil[a]{Department of Astronomy \& Astrophysics, The Pennsylvania State University, 525 Davey Lab, University Park, PA, USA, 16802}
\affil[b]{Center for Exoplanets \& Habitable Worlds, The Pennsylvania State University, University Park, PA, USA, 16802}
\affil[c]{NASA Earth and Space Science Fellow}
\affil[d]{Homer L. Dodge Department of Physics and Astronomy, University of Oklahoma, 440 W. Brooks Street, Norman, OK 73019, USA}
\affil[e]{Department of Physics, Hobart and William Smith Colleges, 300 Pulteney Street, Geneva, NY, 14456, USA}
\affil[f]{Department of Astronomy, Box 351580, University of Washington, Seattle, WA 98195, USA}
\affil[g]{NASA Sagan Fellow}
\affil[h]{Massachusetts Institute of Technology}
\affil[i]{University of California Irvine}

\authorinfo{E-mail: gudmundur@psu.edu}
\begin{document}
\maketitle

\begin{abstract}
The Transiting Exoplanet Survey Satellite (TESS, launched early 2018) is expected to find a multitude of new transiting planet candidates around the nearest and brightest stars. Timely high-precision follow-up observations from the ground are essential in confirming and further characterizing the planet candidates that TESS will find. However, achieving extreme photometric precisions from the ground is challenging, as ground-based telescopes are subject to numerous deleterious atmospheric effects. Beam-shaping diffusers are emerging as a low-cost technology to achieve hitherto unachievable differential photometric precisions from the ground. These diffusers mold the focal plane image of a star into a broad and stable top-hat shape, minimizing photometric errors due to non-uniform pixel response, atmospheric seeing effects, imperfect guiding, and telescope-induced variable aberrations seen in defocusing. In this paper, we expand on our previous work (Stefansson et al. 2017; Stefansson et al. 2018 [submitted]), providing a further detailed discussion of key guidelines when sizing a diffuser for use on a telescope. Furthermore, we present our open source Python package \texttt{iDiffuse} which can calculate the expected PSF size of a diffuser in a telescope system, along with its expected on-sky diffuser-assisted photometric precision for a host star of a given magnitude. We use \texttt{iDiffuse} to show that most ($\sim$80\%) of the planet hosts that TESS will find will be scintillation limited in transit observations from the ground. Although \texttt{iDiffuse} has primarily been developed to plan challenging transit observations using the diffuser on the ARCTIC imager on the ARC 3.5m Telescope at Apache Point observatory, \texttt{iDiffuse} is modular and can be easily extended to calculate the expected diffuser-assisted photometric precisions on other telescopes.
\end{abstract}

\keywords{photometry, diffusers, exoplanets, transits}
\newcommand\blfootnote[1]{%
\begingroup
\renewcommand\thefootnote{}\footnote{#1}%
\addtocounter{footnote}{-1}%
\endgroup
}
\section{Introduction}
The Transiting Exoplanet Survey Satellite (TESS) \cite{ricker2015}, launched in early 2018, will survey the whole sky for transiting planets around the nearest and brightest stars. The predicted planet yield of TESS has been studied by numerous groups (e.g., \cite{sullivan2015,ballard2018,barclay2018}) that predict that TESS will detect a multitude of planets---including hundreds of Neptune-sized planets and dozens of terrestrial Earth-sized planets---orbiting stars that are sufficiently bright for detailed follow-up characterization studies from the ground and with future space-based observatories such as JWST.

While TESS will transform our understanding of the diversity of exoplanet systems around the nearest and brightest stars, ground-based follow-up observations will be critical in confirming the planetary nature of the transiting signals TESS finds, and will play a key role in maximizing the scientific yield of the mission. These ground-based follow-up observations include adaptive-optics and seeing-limited imaging, high precision radial velocity observations, and further ground-based photometry.

In this paper, we restrict our focus on achieving precision photometry from the ground. High precision timely ground-based photometry will play a key role in recovering the ephemerides of a number of planets, which will be essential for efficient scheduling of the most important candidates for further study with JWST\cite{benneke2017,stefansson2018}. Furthermore, high precision ground based photometry can be used as a planet confirmation tool by studying transit depths of planet candidates in different bands \cite{tingley2014}. Such observations can now be performed efficiently by simultaneously observing in a number of different bands with high-precision high-cadence specialized instruments such as MUSCAT\cite{narita2015}, and HiPERCAM\cite{dhillon2016}. Furthermore, high precision photometry can be used to detect transit time variations (TTVs), which can directly give us a handle on the mass of planets in certain planetary system architectures\cite{mazeh2013}. Although high precision photometric observations have been done successfully from space by many groups\cite{beichman2016,benneke2017}, the large number of expected planet candidates from TESS places a great need on the availability of precision photometric instruments from the ground.

However, high precision photometry is difficult to achieve from the ground due to deleterious effects from the atmosphere including differential atmospheric extinction, scintillation, transparency fluctuations and seeing effects. Additionally, inter-pixel sensitivity, telescope guiding effects, and the day-night cycle all affect the photometric precision from the ground. 

Beam-shaping diffusers are emerging as an inexpensive technology to achieve high precision photometry from the ground. Fabricated using specialized nano-fabrication techniques, diffusers are capable of deterministically molding starlight into a broad and stabilized Point Spread Function (PSF) on a telescope imaging array. By spreading out the light over many pixels, allows the observer to increase their exposure times, allowing them to gather more photons and increasing the overall observing efficiency. This increase in observing efficiency ensures that scintillation errors are averaged down, but scintillation errors are the dominating source of error when observing bright stars on large telescopes from the ground \cite{stefansson2017}. These diffusers can be installed on telescopes large and small, both of which play an important role in follow-up efforts from the ground. Smaller telescopes have larger fields of views, allowing them to observe a larger number of reference stars---which becomes especially important when observing bright planet hosts, as nearby bright reference stars are often not available. With their larger diameters, the large telescope will be able to achieve better ultimate photometric precisions due to their enhanced light-collecting area, to both gather more photons, but also better average over scintillation errors.

The use of diffusers for high precision ground-based photometric applications has been discussed in detail in \cite{stefansson2017,stefansson2018}. In these proceedings, we expand the discussion on the adaptation of diffusers in different telescopes, discussiong how to size them to yield a required PSF size. To better plan for diffuser-assisted observations from the ground, we present our open-source code \texttt{iDiffuse}, which can be used to calculate the expected diffuser-assisted photometric precision of a telescope system for given host-star. Although currently mostly designed to calculate the expected photometric precision for the diffuser on the ARCTIC imager \cite{huehnerhoff2016} on the ARC 3.5m \cite{stefansson2017}, \texttt{iDiffuse} can be easily extended to calculate the diffused photometric preicsions of other systems. We use \texttt{iDiffuse} to inform what role diffusers can play in the TESS era. 

This paper is structured as follows. Section \ref{sec:diffusers} gives a short description of diffusers, and how they can be used to achieve high-precision ground-based photometry, giving key best-practices in how to adapt diffusers for use on telescopes for high-precision photometry applications from the ground. Section \ref{sec:planning} describes \texttt{iDiffuse} to better help plan diffuser-assisted follow-up observations of transits. Section \ref{sec:tess} further discusses the role that diffusers can play in the TESS era and beyond. We conclude the paper in Section \ref{sec:summary} with a short summary of the paper.

\section{Transformational Technologies: Diffuser-aided photometry}
\label{sec:diffusers}
Diffusers and their use for high precision photometry application has been further discussed elsewhere \cite{stefansson2017}. In these proceedings, we provide a short description of diffusers here and we refer to \cite{stefansson2017} for a further discussion.

\subsection{Description of diffusers}
As discussed in \cite{stefansson2017}, there are three main types of diffusers including ground-glass diffusers, holographic diffusers and lastly Engineered Diffusers from RPC Photonics\footnote{\url{https://www.rpcphotonics.com/engineered-diffusers-information/}}. We restrict our discussion to Engineered Diffusers from RPC Photonics as these diffusers are currently the only diffusers that can be used to create close-to a top-hat PSF, while the other diffuser types offer either Gaussian or Laurentzian output PSFs.

Engineered Diffusers are made using a laser writing process where a modulated UV laser on a precision XY stage is scanned across a surface to deterministically write the shape and size of the diffuser surface features. These surface features are arranged in individually manipulated unit cells or microlenslets, whose size, shape, and location are varied according to a pre-defined probability distribution to ensure a stable intensity profile and light distribution pattern at the output \cite{rpcpatent}. Further, these unit cells can be carefully designed to avoid discontinuities between the cells to minimize both scattered light and diffraction artifacts at the output \cite{sales2004}.

Engineered Diffusers can be made out of plastic polymers on top of a glass substrate, and can also be etched directly onto a Fused Silica glass substrate. The polymer-on-glass diffusers are inexpensive to make, and can easily be replicated in various sizes, with current capabilities allowing up to $235\unit{mm} \times 235\unit{mm}$ footprints. Polymer-on-glass diffusers are available off-the-shelf with various opening angles (\footnote{\url{https://www.rpcphotonics.com/product-category/polymer-on-glass/}}). These diffusers can work in environments seen in most optical telescope systems, including high fluctuations in humidity and temperature.We have operated the diffuser on the ARC 3.5m telescope for almost 2 years now, which routinely sees large day/night and seasonal temperature changes, with no apparent degradation seen in the diffuser performance.

Although etched Fused-Silica Engineered Diffusers are more expensive (cost driver being the longer fabrication process), these diffusers are useful for near-infrared applications where the glass substrate has to be cooled to cryogenic temperatures. These diffusers have been used successfully on the NIR for precision photometry applications on the WIRC instrument \cite{wilson2003} at Palomar as is discussed in \cite{stefansson2017}.

\subsection{Achieving high-precision ground-based photometry with diffusers}
There are a number of error sources that impact the photometric precision achievable from the ground. We refer the interested reader to \cite{mann2011} and \cite{stefansson2017}, for a more thorough discussion on reaching sub-mmag photometry. Here we quickly discuss how diffusers can be used to reach sub-millimag photometry from the ground.

\paragraph{Inter-pixel Sensitivity:} CCD detector pixels have inherently different sensitivities to light due to QE variations at the few percent level or lower. This effect is in principle corrected for to a high degree by taking good flat-field frames. However, in practice, due to guiding errors, and the intra-pixel sensitivity of the pixels themselves, if the light of a star is only spread over a small number of pixels, this results in photometric noise correlated with guiding. By spreading out the light over a large number of pixels this correlation is minimized, placing less stringent requirements on guiding precision.

\paragraph{Detector Saturation:} By spreading out the light over many pixels, diffusers increase the available dynamic range of  observations, allowing for longer exposure times to gather more photons per exposure. Furthermore, due to the stable and uniform diffused PSF the risk of saturating part of the PSF (i.e., due to seeing variations) is minimized. Although some structure is generally seen in on-sky diffused PSFs, as discussed in \cite{stefansson2017} an even further smoothed PSF can be obtained by rotating the diffuser during an exposure.

\begin{figure}[H]
	\begin{center}
		\includegraphics[width=1.0\columnwidth]{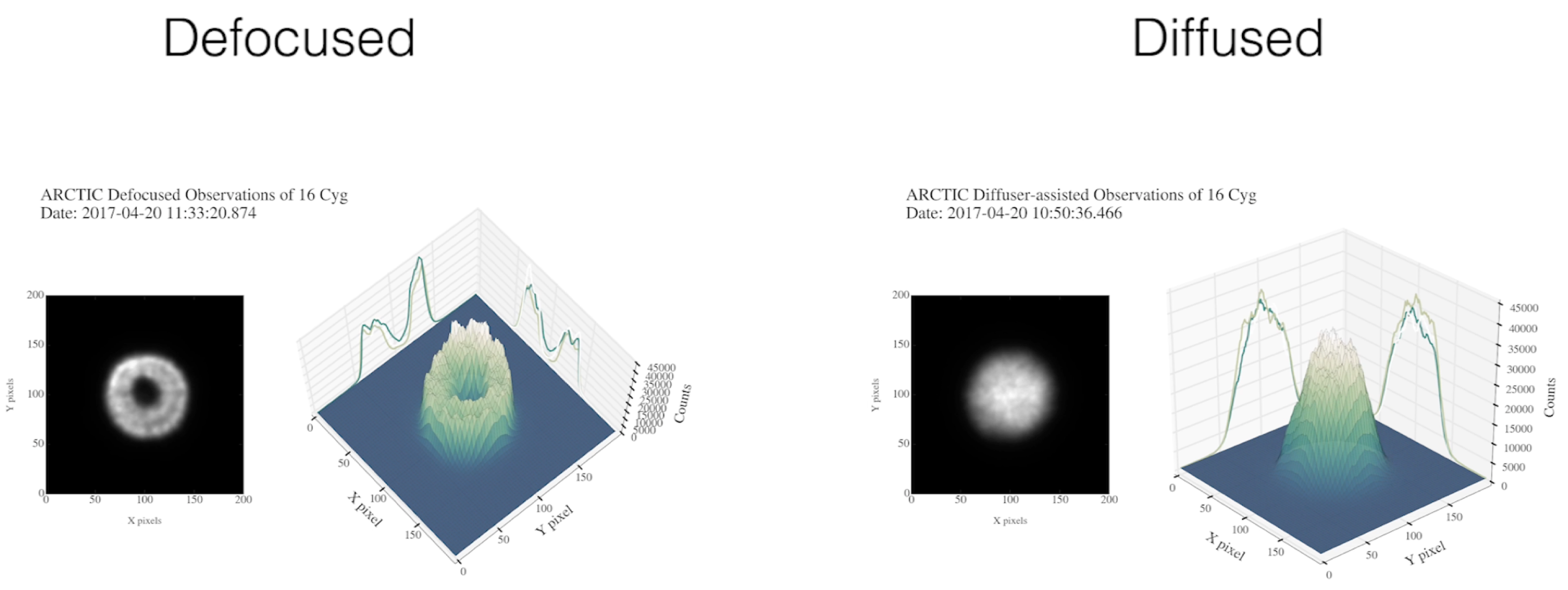}
	\end{center}
	\caption{A 2D/3D comparison of defocused (left) and diffused (right) PSFs on the ARCTIC imager on the ARC 3.5m telescope as a function of time. A video version of this figure is available online (see text). The diffused PSF changes from exposure to exposure, while the diffused PSF is stable throughout the observations.}.
\label{fig:psf}
\end{figure}

\paragraph{Unstable PSF:} Although defocusing has been used to successfully achieve high precision photometry from the ground by many groups in the literature, defocused PSFs are not optimal for precision photometry as they vary with time due to seeing effects and other telescope aberrations. Further, the canonical 'donut PSF' seen in agressive defocusing has little-to-no signal in the central pixels, resulting in a larger photometric aperture than would be achieved with a top-hat PSF. In Figure \ref{fig:psf} we compare diffused and defocused PSFs as a function of time taken with ARCTIC imager on the ARC 3.5m. The observations were 30 minutes each performed on the same star 16 Cyg A on the same night. The defocused observations were done first, at a slightly better airmass than the diffused observations. A video version of this figure is available online\footnote{\url{https://www.youtube.com/watch?v=U6cv1_-qA0o}}. The defocused PSF varies with time causing irregular peaks to appear in the PSF in different locations while the diffused PSF is stable throughout the observations. The stable diffused PSF results in less systematic photometric noise.

\paragraph{Scintillation:} 
Scintillation is caused by starlight being focused inside or outside the telescope primary \cite{dravins1998}, causing intensity variations reducing the achievable photometric precision. Scintillation can be one of the largest source of error in ground-based photometry, especially for the brightest stars. We discuss scintillation further in the context of ground-based follow-up in the TESS era in Section \ref{sec:tess}.

\section{Adapting diffusers to telescope imaging systems}
In this section we list a few key guidelines to think about when adapting a diffuser to a telescope system to produce a given PSF size (Section \ref{sec:install}), and Section \ref{sec:list} gives a list of telescopes where diffusers are being used or considered for high precision photometry applications.

\subsection{Installation and sizing guidelines}
\label{sec:install}
\begin{figure}[H]
\begin{center}
\includegraphics[width=0.8\columnwidth]{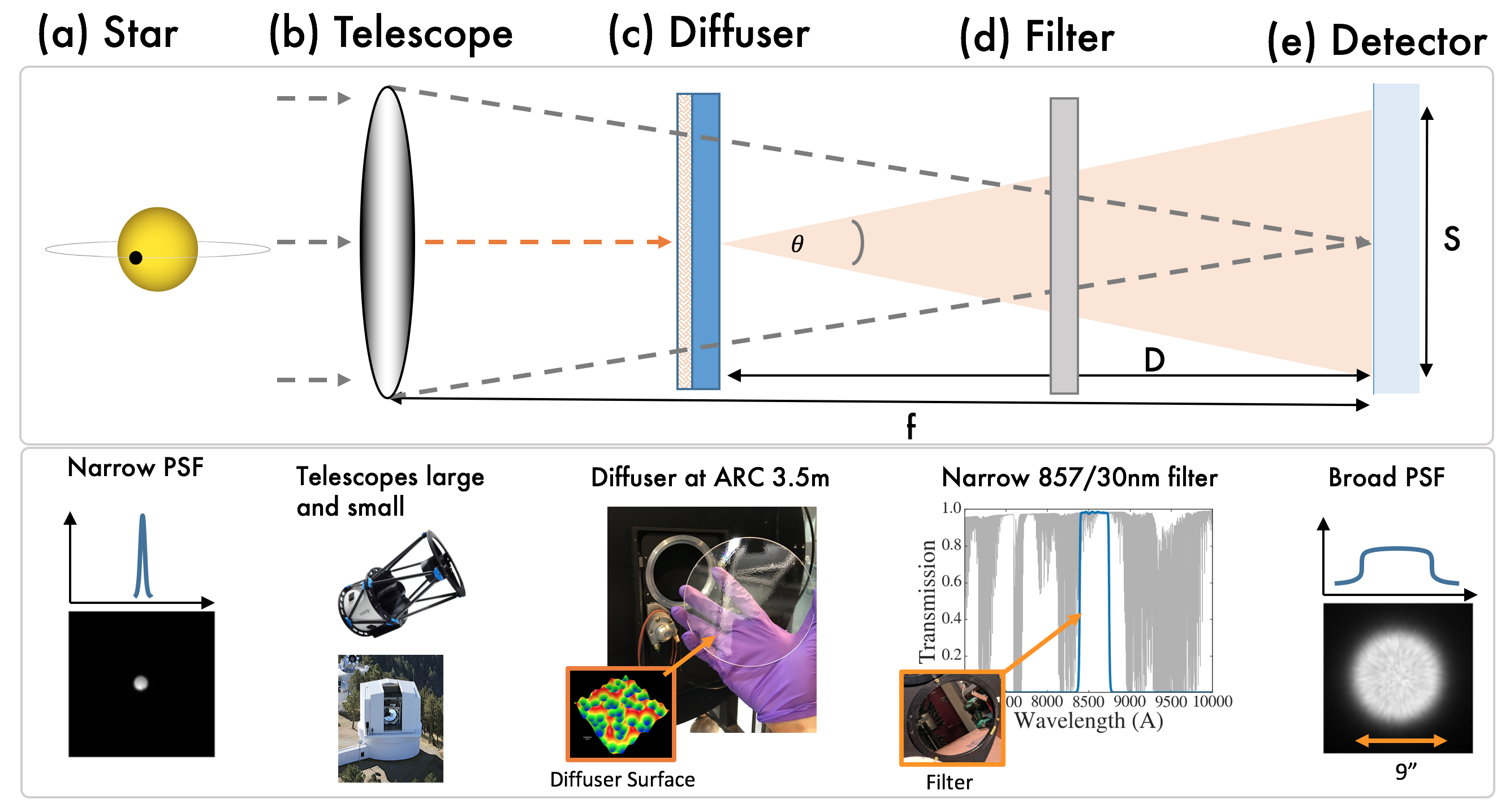}
\end{center}
\caption{Overview of installing a diffuser in a typical converging beam telescope assembly. Light from a star (a) is focused by a telescope (b) through a diffuser (c) and filter (d), onto the telescope imaging array (e), resulting in a broad and stable PSF. The diffuser can equally well be placed in front of or behind the filter of choice. Figure expanded from \cite{stefansson2017}. Diffuser surface image reproduced with permission from RPC Photonics.}
\label{fig:diffuser_path}
\end{figure}

Figure \ref{fig:diffuser_path} gives an overview of a diffuser in a typical converging beam telescope assembly. In \cite{stefansson2017}, we performed lab-tests of diffusers in collimated and converging beams, confirming the output size of the PSF size at Full Width at Half Maximum (FWHM) is given by the following equation,
\begin{equation}
S = \tan(\theta) D,
\label{eq:fwhm}
\end{equation}
where $S$ is the FWHM of the diffused PSF, D is the optical distance of the diffuser from the focal plane (i.e., weighted by the index of refraction $n$), and $\theta$ is the opening angle of the diffuser (see definition in Figure \ref{fig:diffuser_path}). This equation works for simple imaging systems, but if a large number of imaging optics are between the diffuser and the imaging plane, then we recommend validating the expected PSF FWHM from this equation with non-sequential simulations in Zemax. To facilitate the modeling of diffusers in physical systems, RPC Photonics provides the as-measured scattering BSDF files freely online for a variety of different off-the-shelf Engineered Diffusers with different opening angles \footnote{\url{https://www.rpcphotonics.com/bsdf-data-optical-diffusers/}}.

Further points to keep in mind when adapting diffusers for high precision photometry applications are:
\begin{itemize}
\item The diffuser should ideally be placed in a location where the telescope beam footprint is large---e.g., in a collimated beam or in a converging beam away from focus---to illuminate a large number of diffuser the microlenslet-unit cells. Illuminating over a large number of these unit cells better helps average over the distribution of sizes and shapes of the microlenslets, creating a more stable output PSF.
\item Designing and fabricating diffusers with larger opening angles is easier than designing smaller angles, if a top-hat like shape with a steep fall off is desired. Therefore, this places constraints on the absolute distance of the diffuser from the focal plane to not create diffused PSFs that are large, increasing sky-background noise and/or cause stars to blend together.
\item To minimize the effects of seeing changing the diffused PSF, we recommend sizing the diffuser so that the resulting diffused PSF FWHM is least a factor of 3-4 times larger than the median seeing disk. In practice, we have diffused to 9-10\arcsec creating a stable PSF throughout the observations (e.g., Figure \ref{fig:psf}). This usually translates to at least larger than 3-4\arcsec on good observing sights, and larger than 4-6\arcsec on observing sites with poor seeing.
\item Diffusers are most easy to install in a dual filter wheel (see example in Figure \ref{fig:install}). The diffuser can then be placed in one of the filter slot, and easily combined with a filter of choice in the other filter wheel. Although we have not specifically tested this yet, if a dual filter wheel is not available, the diffuser and the filter could potentially be placed on top of each other in a single filter wheel slot. However, we recommend maintaining a small wedged gap between the diffuser and the filter to minimize etaloning. This could be done with a thin 3D printed 'picture-frame' wedge placed between the filter and diffuser, to not impact the clear aperture of the filter and diffuser.

\begin{figure}[H]
\begin{center}
\includegraphics[width=0.9\columnwidth]{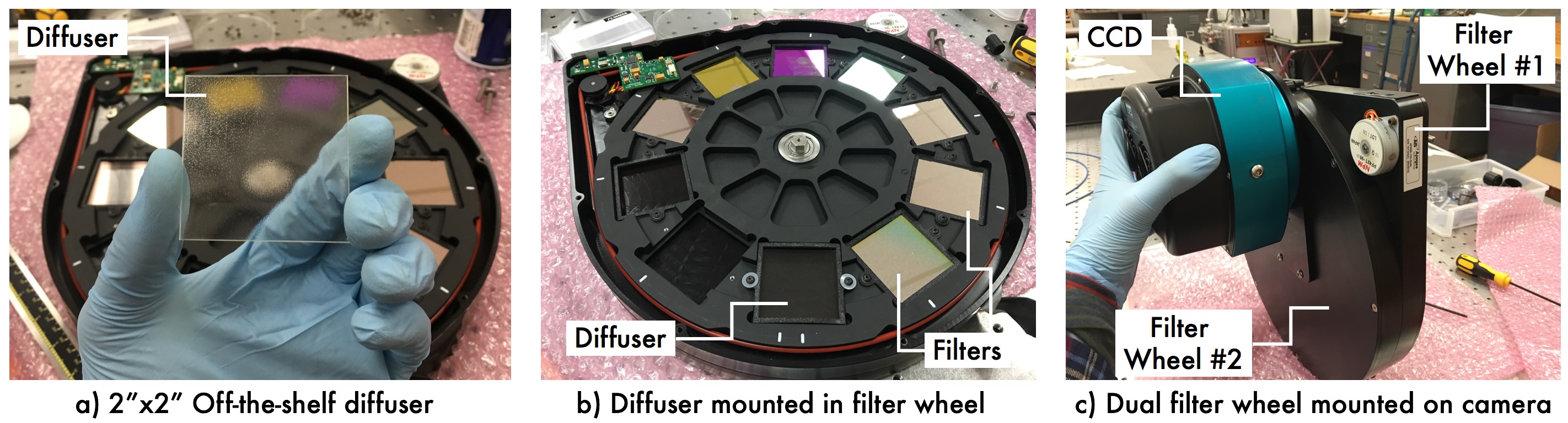}
\end{center}
\caption{Installing a diffuser in a dual filter wheel: a) Example 2"x2" square off-the-shelf Engineered Diffuser from RPC Photoncs. b-c) Diffuser mounted in one of two filters in a dual-filter wheel setup offers an easy way to choose between non-diffused and diffused observations.}
\label{fig:install}
\end{figure}

\item We are currently exploring if a polymer-on-glass diffuser can be placed directly on a filter. This is not practical for interference filters, such as the commonly used SDSS Sloan suite of filters \cite{fukugita1996} as these filters are normally coated on both sides of the glass relying on the air/filter boundary on both ends of the filter to define the bandpass. However, putting a diffuser directly on filtered glass (such as the Johnson-Cousins suite of filters \cite{bessell1990}) might be a possibility, as these filters do not rely on interference patterns on the glass surfaces, but rather the transmission curve is set by the inherent transmission properties of the glass substrate. These filters offers interesting avenues for future work, offering further ease-of-installation in telescope systems.

\item Care should be taken to not have the diffuser overfill any optics downstream of the diffuser, as that will reduce throughput and cause vignetting at the focal plane. This can be minimized by strategic sizing and placement of the diffuser in the telescope beam path.
\end{itemize}

\subsection{A list of diffusers being used in telescopes}
\label{sec:list}
A number of groups have or are considering adapting diffusers to telescopes for precision photometry applications. Table \ref{tab:diffusers} lists some of the telescopes where we know that diffusers are in active use, are currently being tested, and/or are being considered for precision transit photometry.

\begin{table}[H]
\centering
\caption{A list of telescopes and observatories where diffusers have been used, tested, and/or are being explored for high-precision photometry applications. The use of diffusers for precision photometry applications for the first three telescopes is discussed in detail in \cite{stefansson2017}. All diffusers are polymer-on-glass Engineered Diffusers unless otherwise noted.}
\vspace{0.2cm}
{\footnotesize
\begin{tabular}{l l l}
\hline\hline
Telescope, Observatory                       	& Instrument         & Notes                                          \\ \hline
ARC 3.5m, Apache Point Observatory         	& ARCTIC             & In active use                                  \\
Hale 5m, Palomar Observatory                	& WIRC               & Fused Silica diffuser in NIR, in active use    \\
PlaneWave 0.6m, Davey Lab Observatory        	& Aspen CCD          & In active use for research and teaching        \\
Nordic Optical Telescope (2.4m), La Palma    	& ALFOSC             & Tests performed, final diffusers in procurement\\
WIYN 0.9m Telescope, Kitt Peak Observatory   	& Half Degree Imager & In active use                                  \\
0.5m Telescope, Perkin Observatory           	& CCD imager         & In active use for research and teaching        \\
Robotic 30cm Telescope, Las Campanas         	& CCD imager         & Installed in dual filter wheel, in active use  \\
ARCSAT 0.5m, Apache Point Observatory        	& FlareCam           & Diffuser in procurement                        \\
Okayama 1.88m Telescope, Okayama Observatory    & MuSCAT             & Diffusers under development/procurement        \\
MuSCAT 2 Telescope (1.5m), La Palma          	& MuSCAT2            & Diffusers under development/procurement        \\
Gran Telescopio Canarias (10.4m), La Palma   	& HiPERCAM           & Tests proposed, under study                    \\
Central Texas Astronomical Society Telescopes   & Various Imagers    & Tests proposed on small 8-12inch telescopes    \\ 
DEMONEXT 0.5m, Winer Observatory		& FLI CCD 	     & Tests proposed                                 \\
MORC 0.6m Moore Observatory Telescope     	& Apogee CCD         & Tests proposed                                 \\
0.4m Allegheny Observatory Telescope         	& CCD Imager         & Tests proposed                                 \\
\hline
\end{tabular}}
\label{tab:diffusers}
\end{table}

\section{Further systematics control: combining diffusers and narrow-band filters}
As mentioned in our previous work \cite{stefansson2017}, correlated noise can be further suppressed in ground-based transit light curves by using narrow-band filters spanning a wavelength region with minimal telluric absorption lines. In Stefansson et al. 2017\cite{stefansson2017}, we achieved our highest diffuser-assisted photometric precision of 62ppm in 30 minute bins by using an off-the-shelf 30nm-wide narrow-band filter centered at 857nm from Semrock (hereafter Semrock 857/30nm) in observations of 16 Cyg A and B. However, due to the small size of this filter ($2 \times 2 \unit{inch}$), the filter did not fill the full beam footprint in the ARCTIC imager, reducing the FOV from $8.5\arcmin \times 8.5\arcmin$ to $3\arcmin \times 3 \arcmin$.

\begin{figure}[H]
\begin{center}
\includegraphics[width=0.95\columnwidth]{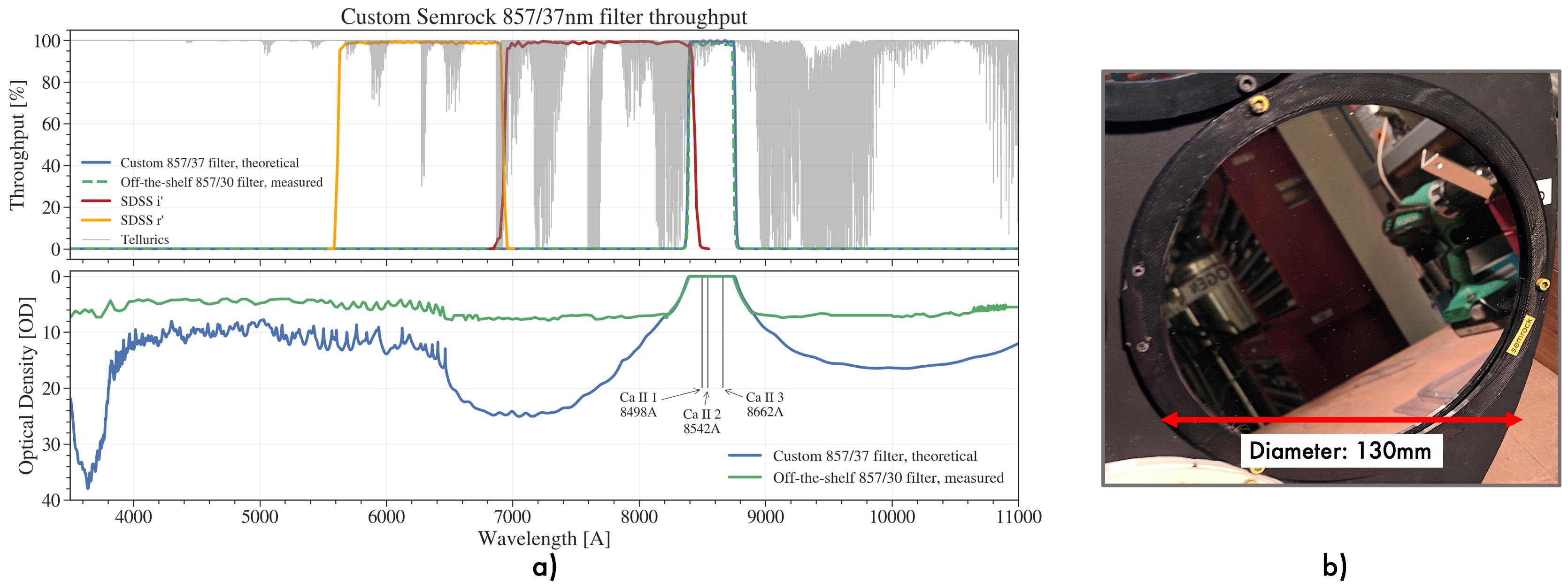}
\vspace{-0.5cm}
\end{center}
\caption{a-top panel) Semrock 857/37nm transmission curve compared to the SDSS $r^\prime$ and SDSS $i^\prime$ transmission curves. The Semrock filter curve is narrower than the broad-band SDSS filters, and has lesser overlap with telluric aborption regions. a-lower panel) Our custom large format 857/37nm transmission curve as compared (theoretical) vs the as-measured transmission curve of the similar 857/30nm off-the-shelf filter. The custom filter has by-design better OD suppression across the full out-of-band region by at least 6 ODs. b) As-delivered large-format custom 857/37nm 130mm Semrock filter in use at the ARC 3.5m telescope.}
\label{fig:semrock}
\end{figure}

To regain the full FOV of ARCTIC, we have worked with AVR Optics \footnote{\url{http://avr-optics.com/}} to make a custom large-format and improved version of this filter (Figure \ref{fig:semrock}). This custom filter was both designed to cover the full FOV of ARCTIC (130mm diameter), while providing enhanced light suppression in out-of-band regions (see Figure \ref{fig:semrock}b) to further minimize out-of-band light-leaks. Specifically, the custom filter increases the Optical Density (OD) suppression from OD$\sim$4 to about OD$\sim$10 from $400\unit{nm}-650\unit{nm}$. This enhanced OD suppression was acquired by a slight increase in the equivalent width of the filter, increasing from 30nm to 37nm.

Of additional interest to stellar activity studies this filter overlaps the Calcium II Infrared-triplet (IRT) (centered at $8498.02 \unit{A}$, $8542.09 \unit{A}$, and $8662.14 \unit{A}$). The core of these lines have their origin in the stellar chromosphere, and have been shown to be excellent tracers of stellar activity in the NIR \cite{robertson2016,sarkis2018}. Although more work remains to characterize the behavior of active stars in this filter, with its minimal overlap with telluric lines this filter offers an observationally inexpensive path to probing stellar magnetic activity.

\section{iDiffuse: calculating expected diffused-assisted photometric precisions}
\label{sec:planning}
To better plan for challenging transit observations with the diffuser on ARCTIC, we have developed a photometric precision package in Python called \texttt{iDiffuse}. The package includes photometric error terms from photon, dark, readout, digitization, sky-brightness, and scintillation noise, along with an additional error-term for the number of reference stars used. \texttt{iDiffuse} is implemented in object oriented Python, based on commonly used community tools in astronomy---including \texttt{astropy} \cite{astropy2013} and the \texttt{pysynphot} the photometry package from STScI \cite{pysynphot2013}\footnote{\url{http://pysynphot.readthedocs.io/en/latest/}}. Although the ensuing discussion focuses primarily on calculating the diffuser-assisted photometric precision of the ARCTIC/diffuser system, \texttt{iDiffuse} is written to be modular and can be easily extend to calculate diffuser-assisted photometric precisions on other telescopes. \texttt{iDiffuse} is freely available on GitHub\footnote{\url{https://gummiks.github.io/idiffuse/}}.

Below we describe our precision calculations in more detail.

\subsection{Calculating expected diffuser-assisted photometric precisions}

\subsubsection{Stellar and sky fluxes}
\begin{figure}[H]
\begin{center}
\includegraphics[width=0.8\columnwidth]{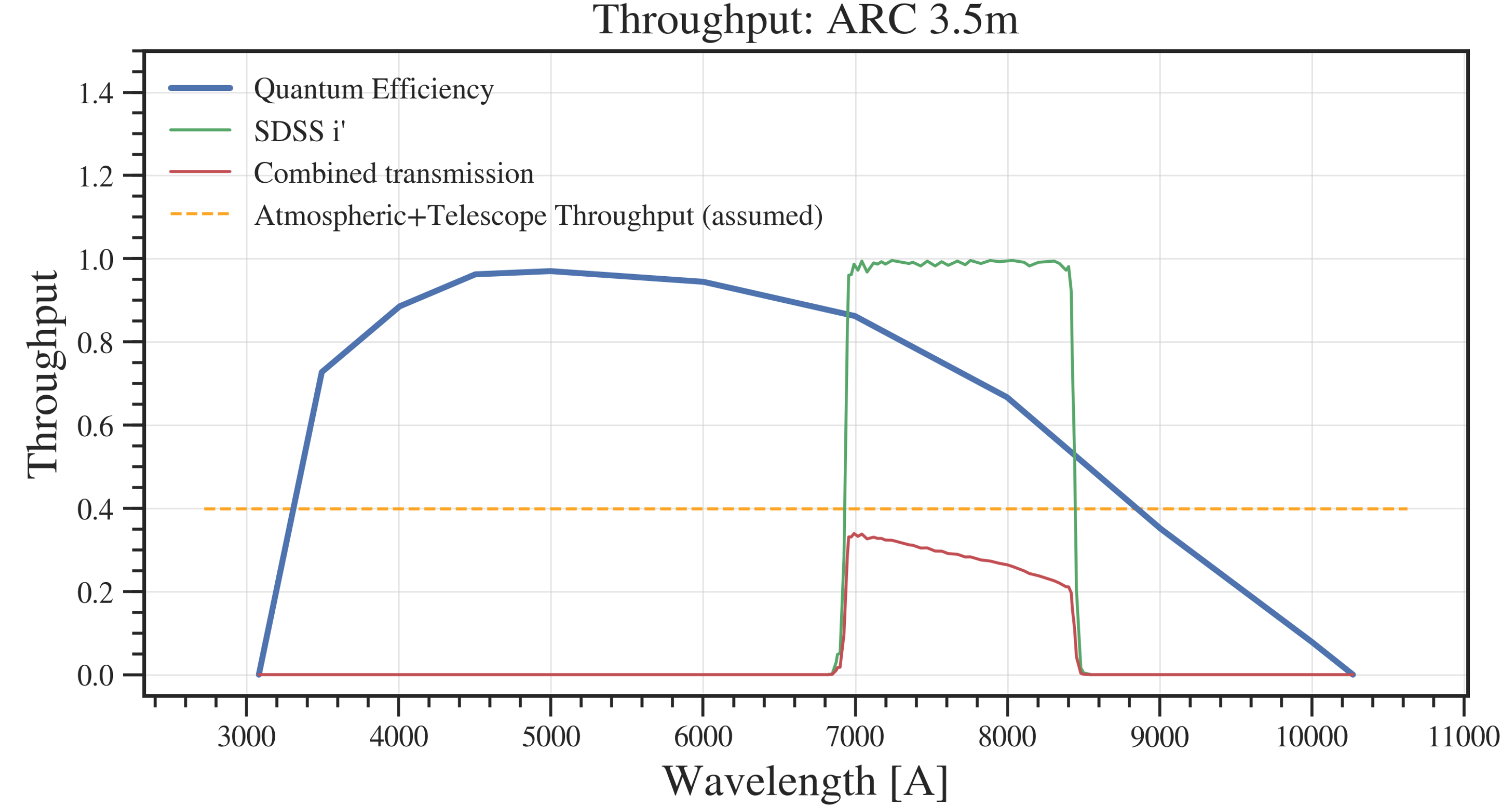}
\vspace{-0.5cm}
\end{center}
\caption{Example throughput plot from \texttt{iDiffuse}. From our on-sky measurements using ARCTIC with the diffuser, we have empirically determined a total atmosphere + system throughput of $\sim$40\% (see orange line). Further work remains in more precisely characterizing this number, and its wavelength dependence.}
\label{fig:throughput}
\end{figure}

To get stellar fluxes incident on the imaging array for a star of a given Vega magnitude, we use the open-source \texttt{pysynphot} package. We refer the reader to the \texttt{pysynphot} documentation on how \texttt{pysynphot} calculates the incident photon flux, but in short \texttt{pysynphot} follows standard formulas to calculate the photon flux from AB magnitudes, which are then renormalized to Vega magnitudes. Further, \texttt{pysynphot} provides an easy-to-use interface to convolve different filters bandpasses, quantum efficiency and transparency curves, making it easy to calculate the resulting stellar flux after accounting for the throughputs of different telescope and atmospheric elements. Although \texttt{pysynphot} was primarily developed to be used with the Hubble space telescope, it is easy to to modify the flux counts to account for the different telescope size.
Although we currently do not have an accurate model of the throughput of the ARC 3.5m telescope, we did our best to incorporate the reflectivity of the different telescope and ARCTIC lens elements including the diffuser to prove a working best-effort throughput model. From our empirical on-sky measurements in the red-optical (the bandpass where we have performed most of our precision photometry), we found that we needed to incorporate an additional flat 50\% throughput factor for the atmosphere to account for the as-measured count rates on the ARCTIC detector, yielding a total of $40\%$ system throughput for the atmosphere and telescope (excluding quantum efficiency) in the red optical (see Figure \ref{fig:throughput}). The nature of this factor and how it varies towards bluer wavelengths is a subject of further study, but the exact throughput number can be easily changed, updated, and made wavelength dependent as more information is available. 

\subsubsection{Photometric precision:}
To calculate the total photometric precision, we follow the steps outlined in \cite{stefansson2017}. To summarize that discussion, we calculate the total photometric errorbars using the following equation,
\begin{equation}
N = \frac{\sqrt{V_* + n_{\mathrm{pix}} \left(1 + \frac{n_{\mathrm{pix}}}{n_b} \right) \left(V_S + V_D + V_R + V_f \right)}}{G},
\label{eq:phot}
\end{equation}
where $G$ is the gain of the CCD in electrons/ADU, $V_*$ is the variance of the net background subtracted counts in the aperture from the star (unit: electrons$^2$), $n_{\textrm{pix}}$ is the number of pixels in the aperture, $n_b$ is the number of pixels used to estimate the mean background sky signal, $V_S$ is the variance in the sky background signal per pixel (unit: electrons$^2$/pixel), $V_D$ is the variance in the dark current signal per pixel (unit: electrons$^2$/pixel), and $V_R$ is the variance of the read noise per pixel (unit: electrons$^2$/pixel), and the last term $V_f$ is the variance in the digitization noise within the A/D converter (unit: electrons$^2$/pixel). The variances in Equation \ref{eq:phot} can be related to the following fluxes (in ADU) and the gain (in $e/ADU$) with the following equations,
\begin{equation}
V_* = |G F_*|, \qquad V_S = |G F_S|, \qquad V_D = |F_D|, \qquad V_R = |F_R^2|, \qquad V_f = |G^2\sigma_f^2|,
\label{eq:variances}
\end{equation}
where $F_*$ is the net background subtracted counts in the aperture from the star in ADUs, $F_S$ is the sky background signal in ADU/pixel, $F_D$ is the dark current signal in electrons/pixel, and $F_R$ is the read noise in electrons/pixel/read, and $\sigma_f$ is an estimate of the 1-$\sigma$ error introduced within the A/D converter with a value of $\sim$0.289ADU.

We then calculate the relative flux error with the following formula,
\begin{equation}
\sigma_{\mathrm{rel\_flux}} = \frac{N}{F} \sqrt{1 + 1/n_E},
\label{eq:relflux}
\end{equation}
where $N$ is the error in ADU as given by Equation \ref{eq:phot}, and $F$ is the total stellar flux in ADU, and $n_E$ is the number of reference stars used. Here we assume that all of the reference stars have the same flux and are uncorrelated. For reference stars with a different flux, Equation in 3 in Stefansson et al. 2017 \cite{stefansson2017} can be used.

We calculate the scintillation term for a given star in units of relative flux as described in Young et al. 1967 \cite{young1967}, and Dravins et al. 1998 \cite{dravins1998}, and further expanded by Osborn et al. 2015 \cite{osborn2015}, with the following approximation:
\begin{equation}
\sigma_{s} = 0.135 D^{-\frac{2}{3}} \chi^{1.75} {\left(2t_{\mathrm{int}}\right)}^{-\frac{1}{2}} e^{\frac{-h}{h_0}} \sqrt{1 + 1/n_{\mathrm{E}}},
\label{eq:stds2}
\end{equation}
where $D$ is the diameter of the telescope in centimeters, $\chi$ is the airmass of the observation, $t_{int}$ is the exposure time in seconds, and $h$ is the altitude of the telescope in meters, and $h_0\simeq8000\unit{m}$ is the atmospheric scale height. The constant 0.135 factor in front has a unit of $\unit{cm^{2/3}s^{1/2}}$, to give the scintillation error in units of relative flux. Following the suggestion by Osborn et al. 2015\cite{osborn2015}, we note that we have multiplied this constant factor by 1.5 from the $0.09 \unit{cm^{2/3}s^{1/2}}$ value originally presented in Young et al. 1967 \cite{young1967} and Dravins et al. 1998 \cite{dravins1998}, to better reflect the median value of scintillation noise. Finally, the $\sqrt{1 + 1/n_{\mathrm{E}}}$ reference star term is derived in Kornilov et al. 2012 \cite{kornilov2012}, describing the number of uncorrelated reference stars $n_\mathrm{E}$. 

Finally, the total photometric error is calculated by,
\begin{equation}
\sigma_{\mathrm{tot}} = \sqrt{ \sigma_{\textrm{rel flux}}^2 + \sigma_{\mathrm{s}}^2}.
\label{eq:stds}
\end{equation}

\subsection{Example usage}
Example usage of \texttt{iDiffuse} can be found in the online documentation: \url{https://gummiks.github.io/idiffuse/}

\subsection{Comparing to on-sky achieved photometric precisions}
Table \ref{tab:expectedvsachieved} compares our achieved on-sky $1 \unit{min}$ diffuser-assisted photometric precision on ARCTIC, to the expected $1 \unit{min}$ photometric precision as calculated with \texttt{iDiffuse}. 
Overall, from Table \ref{tab:expectedvsachieved}, we see that the expected photometric precision as calculated by \texttt{iDiffuse} agrees well to the on-sky photometric precision, with the on-sky precision being generally slightly higher than the expected precision. This discrepancy can be due to many factors, including but not limited to stellar variability in the target-and reference stars, transparency fluctuations due to passing clouds, differential extinction effects etc.

We do note that all of the comparisons in Table \ref{tab:expectedvsachieved} are in the red-optical (SDSS $r^\prime$, SDSS $i^\prime$, and Semrock 857/30nm) as those have been our filters of choice for our precision diffuser-assisted photometry. Therefore, the validity of \texttt{iDiffuse} towards the bluer wavelengths will benefit from on-sky cross-checks at bluer wavelengths. We leave that as subject for further work, and will continue to update \texttt{iDiffuse} as new information becomes available. 

\begin{table}[H]
\centering
\caption{A comparison between on-sky achieved diffuser-assisted photometric observations performed with ARCTIC compared to the precision expected from \texttt{iDiffuse}. The $\sigma_{\mathrm{1min,expected}}$ and $\sigma_{\mathrm{1min,avhieved}}$ are Gaussian-noise scaled to a cadence of $1\unit{min}$ from the unbinned observations.}
\vspace{0.2cm}
{\footnotesize
\begin{tabular}{l c c c c c l}
\hline \hline
Object     & $\sigma_{\mathrm{1min,expected}}$   & $\sigma_{\mathrm{1min,achieved}}$   & Filter           & $t_{\mathrm{exp}}$ & SDSS $i^\prime$ & Notes                                                   \\
-          & [ppm]                               & [ppm]                               & -                & [s]                & [mag]           & -                                                       \\ \hline
TRES-3b    & 404                                 & 541                                 & SDSS $i^\prime$  & 30                 & 11.2            & Transit presented in \cite{stefansson2017}              \\
WASP 85Ab  & 434                                 & 626                                 & SDSS $r^\prime$  & 6                  & 10.0            & Transit presented in \cite{stefansson2017}              \\
16 Cyg A/B & 338                                 & 300                                 & Semrock 857/30nm & 16                 & 5.17            & Flat-line observation \cite{stefansson2017}             \\
K2-28b     & 1261                                & 1532                                & SDSS $i^\prime$  & 32                 & 13.9            & Transit presented in \cite{stefansson2018}              \\
K2-100b    & 305                                 & 358                                 & SDSS $i^\prime$  & 16                 & 10.4            & Transit presented in \cite{stefansson2018}              \\ \hline
\end{tabular}}
\label{tab:expectedvsachieved}
\end{table}

\section{Diffuser-assisted Photometry in the TESS era}
\label{sec:tess}
\begin{figure}[H]
\begin{center}
\includegraphics[width=0.8\columnwidth]{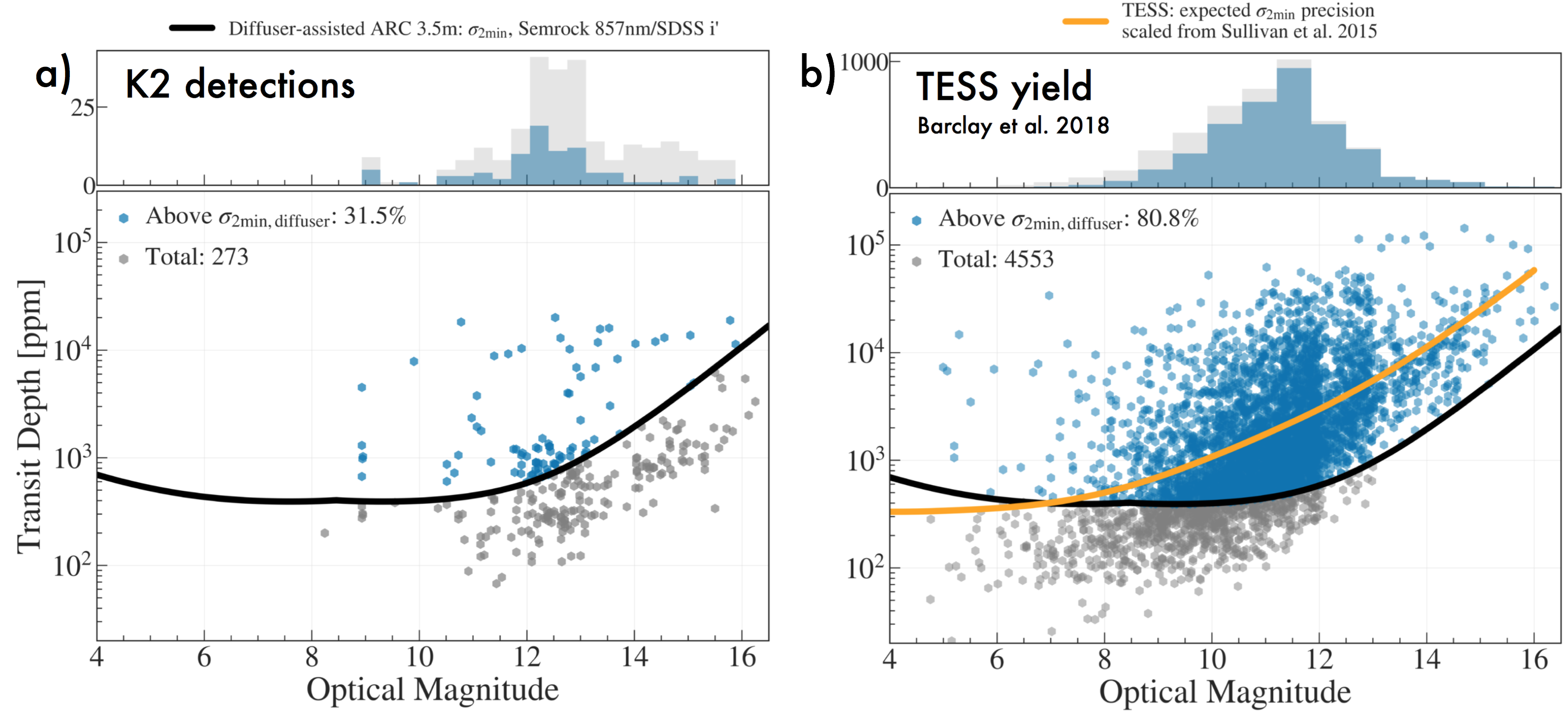}
\end{center}
\caption{Expected diffuser-assisted precision with an ARCTIC+diffuser-like system on a generic 3.5m telescope (with similar pefromance as ARC 3.5m) as a function of red-optical/TESS magnitude showing the expected 2-minute photometric precision ($\sigma_{\mathrm{2min}}$) compared with: a) transiting planets detected by the \textit{K2}-mission; b) expected transiting planet yield from TESS from Barcaly et al. 2018\cite{barclay2018}. We color planets with transits that can be reliably followed up with such a system in blue (above the $\sigma_{\mathrm{2min}}$ curve), and other planet transits in gray. The blue histograms in the upper panels show the distribution of the blue points, and the gray histograms shows the histogram of the respective total planet populations shown in panels a) and b). Additionally shown in b) is the expected $\sigma_{\mathrm{2min}}$ precision of TESS. The black curve assumes observations performed in either SDSS $i^\prime$ or Semrock 857/37nm at an airmass of 1.5 using 1 equally bright reference star to the target star.}
\label{fig:TESS}
\end{figure}

To demonstrate the precision capabilities of diffuser-assisted photometry in the $K2$ and TESS era, in Figure \ref{fig:TESS} we use \texttt{iDiffuse} to calculate the fraction of planet transits an ARCTIC+diffuser-like system on a 3.5m telescope could recover from A) the currently confirmed $K2$ planet sample as of July 2018\footnote{Derived from the Exoplanet Archive \url{https://exoplanetarchive.ipac.caltech.edu/}}, and B) the expected TESS yield from Barclay et al. 2018 \cite{barclay2018}.

The black curves in Figure \ref{fig:TESS} show the expected 2-minute binned precision of an ARCTIC+diffuser like system on a generic 3.5m ground-based telescope capable of observing the target planet host star at an airmass of 1.5. We use this curve as the dividing line between planet transits that we believe such as system will be able to reliably recover planet transits with high $>$$4\sigma$ confidence (above the $\sigma_{\mathrm{2min}}$ curve; blue dots), and low confidence (below the $\sigma_{\mathrm{2min}}$ curve; gray dots). The blue histograms in the upper panels in Figure \ref{fig:TESS} show the distribution of the blue points, and the gray histograms show the distribution of the full corresponding planet sample. This comparison shows that only $30\%$ of the planet transits from $K2$ can be reliably followed up with such as system, while this number increases to $80\%$ for the expected TESS yield. For this calculation, the black curves assumes observations performed at an airmass of 1.5 using an on-chip binning mode of 2x2 with a read time of 5s, no read and dark noise, and assumes the availability of 1 equally bright reference star. We acknowledge that the assumption of the availability of an equally bright reference star below optical magnitudes of Mag$\sim$8-10 starts becoming less likely---and will be highly dependent on the field in question for the brightest stars. The black curves in Figure \ref{fig:TESS} assume observations performed in two filters: either the SDSS i' filter \cite{fukugita1996} or the Semrock 857/37nm filter, whichever is more precise at a given magnitude, resulting in a slight kink in the expected precision-curve at about a magnitude of $\sim$9. Being broader, the SDSS $i^\prime$ filter offers enhanced precision for fainter stars while the Semrock 857/30nm filter---being almost four times as narrow as the SDSS $i^\prime$ filter---offers enhanced precisions for the brighter planet hosts while also being less susceptible to telluric-induced systematics. Additionally shown in Figure \ref{fig:TESS}b is the expected $\sigma_{\mathrm{2min}}$ of TESS from \cite{sullivan2015}, demonstrating that across a wide range of optical magnitudes ($>7 \unit{mag}$), the ARCTIC+diffuser system will be more precise than TESS for a given time bin.

\begin{figure}[H]
\begin{center}
\includegraphics[width=0.8\columnwidth]{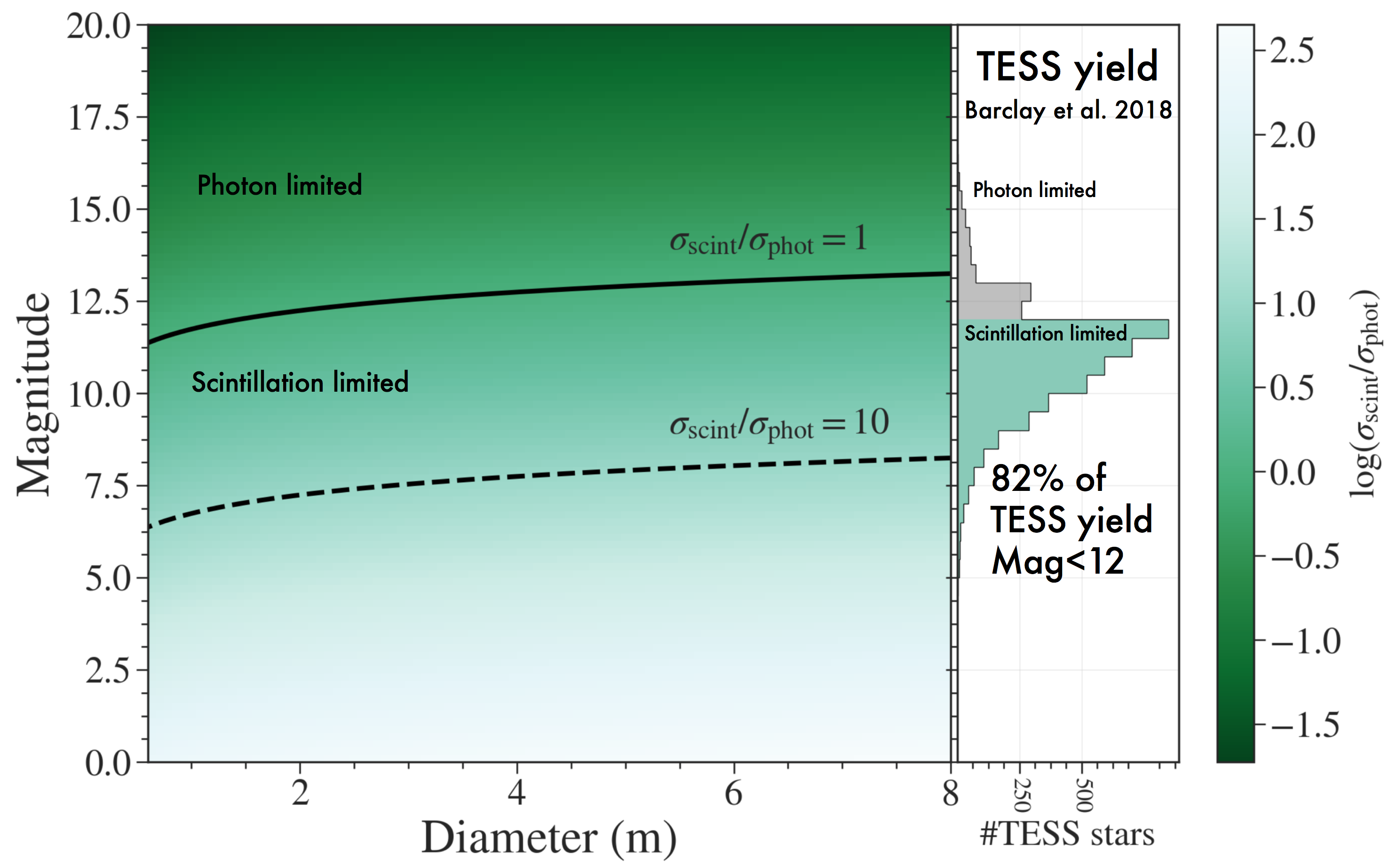}
\vspace{-0.5cm}
\end{center}
\caption{A comparison of photometric and scintillation errors in the stellar magnitude-telescope diameter plane. The solid black curve shows where the photon noise errors equal the scintillation errors, and the black dashed curve shows where the scintillation noise is an order of magnitude larger than the photon noise error. This plot uses \texttt{iDiffuse} assuming an 'ideal telescope' with a flat transmission of $40 \%$, binning mode of 2x2, no readout time, with observations performed at an airmass of 1, and using 1 equally bright reference star. The right panel shows the expected TESS exoplanet host yield, showing that over $80\%$ of the TESS planet hosts will be scintillation limited ($\mathrm{mag} < 12$) in photometric observations from the ground.}
\label{fig:errors}
\end{figure}

In Figure \ref{fig:errors} we use \texttt{iDiffuse} to expand on a Figure we initially presented in \cite{stefansson2017}, where we compare the expected photon noise to the expected scintillation noise in the stellar magnitude-telescope diameter plane. We calculate the expected photometric noise contributions using \texttt{iDiffuse} assuming a perfect telescope (observation performed at airmass 1, a flat transmission of $40 \%$, no readout time, binning of 2x2, altitude of 2,200 \unit{m}) and assume 1 equally bright reference star to the target star. A more realistic observation than this ideal case (i.e., at lower airmasses with a finite readout time) will shift the lines further towards the upper part of Figure \ref{fig:errors}, where scintillation noise fractionally increases. The solid black curve shows where the photon noise $\sigma_{\mathrm{phot}}$ equals the scintillation errors $\sigma_{\mathrm{scint}}$, and the dashed line shows where the scintillation error is an order of magnitude larger than the photon noise. We see from these curves that the dependency on the telescope diameter is a slowly varying function of the diameter $\sim$$D^{-1/6}$.

In the right panel in Figure \ref{fig:errors} we use the $\sigma_{\mathrm{phot}}/\sigma_{\mathrm{scint}} = 1$ curve (solid black curve) as a dividing line to say that $\sim$80\% of the TESS yield will be scintillation limited from the ground across different telescopes. Therefore, to reach the highest photometric precisions in ground-based follow-up transit observations of most of the expected TESS planet hosts, scintillation noise will need to be minimized. 

Scintillation can be minimized in a few ways. First, scintillation can be minimized with high duty-cycle observations, which can be achieved with fast readout detectors with minimum read-noise. Second, scintillation can further be minimized by performing observations on larger telescopes. However, large telescopes generally have smaller fields of view, limiting their access to equally bright reference stars for bright-star transit observations. Third, new dedicated technologies such as conjugate-plane scintillation correctors have been proposed in the literature and tested on sky \cite{osborn2011,osborn2015} to correct and/or minimize scintillation. Although a promising technology, these correctors are currently not widespread due to the need of a dedicated optical setup. Nevertheless, some of the upcoming high precision imagers have been specifically designed with such scintillation correctors in mind (e.g., HiPERCAM \cite{dhillon2016}) to reach the highest photometric precisions.

\section{Summary and future work}
\label{sec:summary}
In this paper we have described diffusers and their use for precision ground-based photometry in the TESS era.
We further summarize our key points in the list below.

\begin{itemize}
\item Diffusers are being incorporated into a number of different observatories.
\item We presented \texttt{iDiffuse}, an open source tool to calculate expected on-sky diffuser-assisted photometric precisions. Although \texttt{iDiffuse} has been written with the diffuser on ARCTIC, \texttt{iDiffuse} is modular and can be easily extended to calculate the diffuser-assisted photometric precision on othere telescopes.
\item Using \texttt{iDiffuse}, we predict that an ARCTIC+diffuser-like system on a 3.5m telescope should be able to recover at high confidence 4 out of every 5 planet transits that TESS will find.
\item Most TESS planet hosts will be scintillation limited in transit observations from the ground. Diffusers help minimize scintillation in conventional telescope imagers by allowing increased duty cycles.
\item We expect that the highest ground-based diffuser-assisted photometric precision will be achieved by adapting diffusers on large telescopes with rapid-readout imagers and scintillation correctors.
\end{itemize}

\acknowledgments
We gratefully acknowledge the work and assistance of Tasso Sales and Laura Weller-Brophy at RPC Photonics, without whose help this project would not have been possible. We thank the observing staff at Apache Point Observatory for all their help with the observations presented here, with special thanks to Jack Dembicky, Candace Gray, William Ketzeback, Russet McMillan and Theodore Rudyk.

This work was directly seeded and supported by a Scialog grant from the Research Corporation for Science Advancement (Rescorp) to SM, LH, JW. This work was partially supported by funding from the Center for Exoplanets and Habitable Worlds. The Center for Exoplanets and Habitable Worlds is supported by the Pennsylvania State University, the Eberly College of Science, and the Pennsylvania Space Grant Consortium. GKS wishes to acknowledge support from NASA Headquarters under the NASA Earth and Space Science Fellowship Program-Grant NNX16AO28H.

These results are based on observations obtained with the Apache Point Observatory 3.5-meter telescope which is owned and operated by the Astrophysical Research Consortium. We acknowledge support from NSF grants AST-1006676, AST-1126413, AST-1310885, AST-1517592, the NASA Astrobiology Institute (NAI; NNA09DA76A), and PSARC. This research made use of the NASA Exoplanet Archive, which is operated by the California Institute of Technology, under contract with the National Aeronautics and Space Administration under the Exoplanet Exploration Program.

Software: 
\texttt{AstroImageJ} \cite{collins2017},
\texttt{Astropy} \cite{astropy2013},
\texttt{Astroquery} \cite{ginsburg2016},
\texttt{batman} \cite{kreidberg2015}, 
\texttt{iDiffuse} (this work),
\texttt{Jupyter} \cite{jupyter2016},
\texttt{matplotlib} \cite{hunter2007},
\texttt{numpy} \cite{vanderwalt2011},
\texttt{pandas} \cite{pandas2010},
\texttt{pysynphot} \cite{pysynphot2013},
\texttt{TelFit} \cite{gullikson2014},

\bibliography{references}
\bibliographystyle{spiebib}

\end{document}